\documentclass[english]{article}
\usepackage[T1]{fontenc}

\usepackage[utf8]{inputenc}
\usepackage[english]{babel}
\usepackage{geometry}
\usepackage{float}
\usepackage{amsmath}
\usepackage{graphicx}
\usepackage{natbib}
\usepackage{lipsum}
\usepackage{natbib}
\usepackage{graphicx}

\makeatletter

\floatstyle{ruled}
\newfloat{algorithm}{tbp}{loa}
\providecommand{\algorithmname}{Algorithm}
\floatname{algorithm}{\protect\algorithmname}

\newcommand{\lyxaddress}[1]{
	\par {\raggedright #1
	\vspace{1.4em}
	\noindent\par}
}

\makeatother

\usepackage{babel}
\begin{document}

\title{Multi-MeV electron occurrence and lifetimes in the outer radiation belt and slot region during the maximum of solar cycle 22}

\author{{R. T. Desai$^{1,2,3}$\thanks{corresponding author: ravindra.desai@warwick.ac.uk}} , J. Perrin$^{1}$, N. P. Meredith$^{2}$, S. A. Glauert$^{2}$, S. Ruparelia$^{1}$, W. R. Johnston$^{4}$}

\date{}
\maketitle

\vspace{-2em}
\lyxaddress{\begin{center}
$^1$Centre for Fusion, Space \& Astrophysics, University of Warwick, UK
\par\end{center}}
\vspace{-2em}
\lyxaddress{\begin{center}
$^2$British Antarctic Survey, Cambridge, UK
\par\end{center}}
\vspace{-2em}
\lyxaddress{\begin{center}
$^3$Blackett Laboratory, Imperial College London, London, UK
\par\end{center}}

\begin{abstract}
The Combined Release and Radiation Effects Satellite (CRRES) observed the response of the Van Allen radiation belts to peak solar activity within solar cycle 22. This study analyses {relativistic and ultra-relativistic electron} occurrence and loss timescales within the CRRES High Energy Electron Fluxometer (HEEF) dataset, including during several strong and severe geomagnetic storms that {all, remarkably,} flooded the slot region with multi-MeV electrons. These allow the first definitive multi-MeV electron lifetimes to be calculated in this region and indicate an elevated risk to satellites in slot region orbits {during periods of heightened solar activity}. The HEEF outer belt loss timescales are broadly in agreement with those from later solar cycles, but differences include longer-lasting sub-MeV electrons near the inner region of the outer belt and faster-decaying multi-MeV electrons near geosynchronous orbit. These differences are associated with higher levels of geomagnetic activity, a phenomenon that enables the spread in the results to be parameterised accordingly. The timescales generally appear well-bounded by Kp-dependent theoretical predictions, but the variability within the spread is not always well-ordered by geomagnetic activity. This suggests the limitations of using pitch-angle diffusion to account for the decay of elevated electrons following geomagnetic storms, and the need for more sophisticated space weather indices for radiation belt forecasting.
\end{abstract}

\section{Introduction}

[1] The Van Allen electron radiation belts consist of two concentric tori encircling the Earth with energies extending up to several MeV \citep{Vanallen59}. Satellites are designed to mitigate or avoid the impacts of these ``killer'' electrons which can lead to the triggering of phantom commands and component failure \citep[e.g.][]{Baker13,Hands18}. The precipitation of energetic particles also directly influences the ionisation and conductivities of the upper atmosphere, and its coupling back to the geospace environment \citep[e.g.][]{Roble77,Orsolini05}. The dynamics of trapped relativistic electron populations are thus a primary focus of space weather research {and forecasting efforts}. 

[2] The electron radiation belts typically exhibit a two-belt structure, as commonly ordered by the \citet{Roederer70} L* parameter. A stable so-called inner belt peaks between L* $\approx$ 1--2 \citep{Selesnick07} and a highly variable outer belt extends between L* $\approx$ 3--8 \citep{Glauert18}. This canonical picture can however rapidly change during geomagnetic storms, with additional peaks in the relativistic particles flux, i.e. new radiation belts, forming at various distances to the Earth, including inside the slot region.  
The first well-known example occurred on 24 March 1991, where a large interplanetary shock near-instantaneously injected a new ultra-relativistic belt extending up to over 50 MeV inside the slot region \citep{Blake1992}. {Elevated energetic particles fluxes during this March 1991 event resulted in two satellite outages, and single event upsets (SEUs) and solar cell degradation on orbiting satellites \citep{Mullen91, Ishikawa2013, Shea93}}. A further significant event occurred during solar cycle 23 following the Hallowe'en storms of 2003 where the outer belt reformed inside the slot region and into the inner belt as a result of enhanced convection and erosion of the plasmasphere \citep{Baker04}. {This similarly caused a satellite outage and extensive anomalies \citep{cho2005failure,noaa2004intense}.} This again occurred in solar cycle 25, when the Mother's/Gannon Day storm of May 2024 resulted in elevated fluxes forming inside the slot region \citep{Pierrard24,Li25}. The reordering of the outer belt into multiple distinct outer belts is also commonly occurring following geomagnetic storms \citep[e.g.][]{Pinto18,Vampola71,Baker13,Chen24}. 

[3] Elevated electron fluxes in the inner magnetosphere, following geomagnetic storms, are observed to decay exponentially across timescales of days-weeks, a phenomenon associated with pitch-angle scattering into the atmospheric loss cone \citep{Meredith06,Meredith09,Baker04,Claudepierre20a}. This process provides a direct method to test diffusion theory \citep{Lyons74,Kennel85,Glauert05,Albert05} and its application within 3-d radiation belt models  \citep[e.g.][]{Fok01,Shprits09,Su10,Reeves12,Glauert14}. As a general overview the following plasma waves have found to be significant.
Within the slot region, the dominant modes have been found to be plasmaspheric hiss \citep{Lyons72,Meredith07}, lightning-generated whistlers  \citep[e.g.][]{Lauben01} and magnetosonic waves \citep{Ma16,Wong22}. Further out, hiss continues to play an important role for electron loss in the outer radiation belt both within the plasmasphere \citep{Summers04,Meredith06} and in plasmaspheric plumes \citep{Summers08}. In addition, electromagnetic ion cyclotron (EMIC) \citep{Miyoshi08,Ross21} and whistler-mode chorus waves \citep{Shprits07,Wang19,Wang23} and outward radial diffusion to the magnetopause \citep{Shprits07radial,Mann16} may also contribute to electron loss in the outer radiation belt. Anthopogenic VLF whistlers and Coulomb collisions have also been found to be effective in the inner belt and slot region \citep{Ripoll14,Cunningham18,Ross19}.

[4] Multi-MeV electron loss timescales have been empirically calculated within a few studies but the energy discrimination has often been limited to a single energy channel \citep[e.g.][]{West81, Seki05,Ripoll15,Baker07}. The Van Allen Probe electron lifetimes reported by \citet{Claudepierre20a} addressed this, determining decay rates within six differential energy channels extending from 1 -- 4 MeV. This study provides analysis of multi-MeV electrons observed by CRRES during {the maximum of} solar cycle 22, providing decay rates within six energy channels up to 4.55 MeV. Moreover, during the VAP mission multi-MeV electrons weren't seen inside L*=2.8 \citep{Baker14} whereas in this study we find that CRRES regularly observed multi-MeV electrons inside the slot region. This enables the first calculation of well-resolved lifetimes in the slot region. To investigate the variability of the loss timescales we also parameterise the results by geomagnetic activity to perform an activity-dependent comparison of empirical electron loss timescales with predictions by pitch angle diffusion theory. 

[5] The paper is structured as follows. In Section \ref{Section2} we outline the concept of pitch angle diffusion and the approach used to calculate the theoretical decay rates. In Section \ref{Section3} we introduce the CRRES instrument and dataset and the automated algorithm used to derive the empirical loss timescales.  In Section \ref{Section4} we then present the {geomagnetic activity-dependent} lifetime analysis of multi-MeV electrons during the severe geomagnetic storms that flooded the slot region and the automated retrieval of lifetimes through the entire dataset. Section \ref{Section5} then contrasts these with the theoretical solutions and previously reported loss timescales. Section \ref{Section6} then summmarises the results of the study. 

\begin{figure}[ht]
     \begin{center}
             \includegraphics[width=1.0\textwidth]{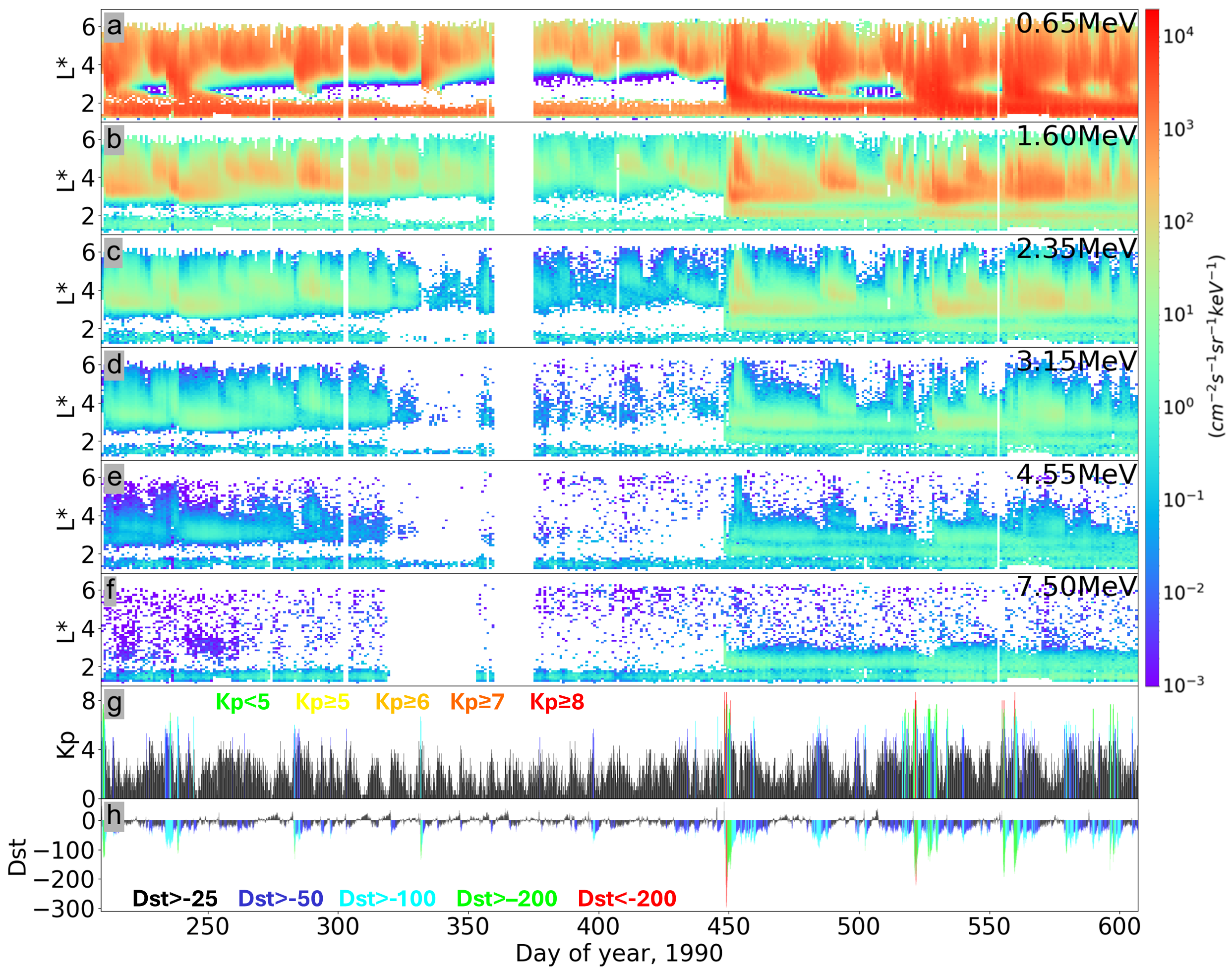}
     \end{center}
     \caption{HEEF daily averaged differential number flux at each second energy channels outlined in Table \ref{table} in subplots a-f throughout the CRRES mission. Subplot (g) and (h) show the respective corresponding Kp and Dst indices.}
      \label{spectrogram}
 \end{figure}

\section{Pitch Angle Diffusion Theory}
\label{Section2}

Pitch angle diffusion is caused by cyclotron and landau resonances. Under the assumption of quasi-linear theory of timescales longer than a drift-period, the pure pitch angle diffusion equation for particle flux, $f$, as a function of time, $t$, and equatorial pitch angle, $\alpha_0$, can be written as 
\begin{equation}
\frac{\partial f}{\partial t} = \frac{1}{T \sin(2\alpha_0)} \frac{\partial}{\partial \alpha_0} \bigg|_{L^*, E} \left(T \sin(2\alpha_0) D_{\alpha\alpha} \frac{\partial f}{\partial \alpha_0}\right),
\label{diffusion}
\end{equation}
where $D_{\alpha\alpha}$ represent pitch angle diffusion coefficients and T($\alpha_0$) = 1.3 - 0.56 sin$\alpha_0$ as the latitude dependence of the bounce period. {We can assume a basis of solutions given by these eigenmodes} {to be separable as  f($\alpha$, t) = A($\alpha$)f(t). {The distribution is therefore represented by a superposition of exponentially decaying orthonormal eigen-states \citep{OBrien2008} and following \citet[][]{Lyons72}, can be assumed to have}},
\begin{equation}
f(t) = Ae^{(-t/\tau)},
\label{decay}
\end{equation}
for an equatorially mirroring pitch angle distribution $A(\alpha_0)$ and decay constant, $\tau$. This leaves
\begin{equation}
\frac{d}{d\alpha_0} \left( D_{\alpha\alpha} T \sin(2\alpha_0) \frac{dA}{d\alpha_0} \right) + \frac{T \sin(2\alpha_0)}{\tau} A = 0,
\end{equation}
a second order difference equation that can be solved as a boundary value problem \citep{ALbert94,Meredith06} for a given pitch angle distribution and as a function of chosen diffusion coefficients. 
The eigenvalues of the operator specify the time scales of the diffusion process \citep[e.g.][]{Albert09,Schulz1974} {which Sturm-Liouville theory assumptions dictate are real and ordered \citep{OBrien2008,Claudepierre22}}. The lowest order eigenvalues therefore yield increasingly longer timescales and these have appeared dominant within observations reported so far at both equatorial and high latitudes {in the outer radiation belt} \citep{Meredith06,Baker07,Claudepierre20a}. However, this may not always be the case if the timescales of the diffusion process are longer than the timescales acting and the higher-order eigenvalues act across similar timescales to those examined {as in the inner slot region and inner radiation belt \citep[][]{Meredith09,Broll23}}. 
The exponentially decaying solution in Equation 2, allows empirical e-folding decays to be derived from observational data, as described in the next section, to serve as a direct comparison to these theoretical solutions.

In this study we utilise the Kp-dependent theoretical approach to Equation 3 of \citet{Glauert24} which accounts for plasmaspheric hiss, upper/lower band chorus waves, EMIC waves, VLF transmitters, lightning‐generated whistlers (LGW) and magnetosonic waves, derived within the \citet{Olson} field model.  \citet{Glauert24} used this approach to report loss timescales up to 1 MeV and at the single multi-MeV channel 2.6 MeV. In this study, we use this approach to show lifetimes within each of the six HEEF energy channels from 0.65 to 4.55 MeV.  The hiss diffusion coefficients are derived from the wave model of \citet{Meredith18} from 2$ \leq L* \leq 6$ using Dynamics Explorer, Double Star, Cluster, THEMIS, and VAP observations, and the PADIE code \citep{Glauert05}. The chorus diffusion coefficients are derived using the observations reported by \citet{Meredith20} for 2$ \leq L* \leq 10$ based upon Dynamic Explorer, Double Star , THEMIS, and VAP, and the PADIE code. The VLF transmitter diffusion coefficients were derived by \citet{Ross18} for $L*<3$ using VAP observations. The EMIC diffusion coefficients were derived by \citet{Ross21} and \citet{Ross20} from 3.25$ \leq L* \leq 7$ using VAP observations. The LGW diffusion coefficients were calculated using PADIE for 2$ \leq L \leq 3$ based on the \citet{Green20} VAP wave model and the densities of \citet{Ozhogin12} from the IMAGE spacecraft. The magnetosonic diffusion coefficients are derived from \citet{Wong22} using VAP observations. Pitch angle diffusion due to collisions with the atmosphere inside the loss cone is also incorporated as described in \citet{Selesnick16} for a tilted dipolar magnetic field with specified eccentricity.

\begin{figure}[ht]
     \begin{center}
             \includegraphics[width=1.0\textwidth]{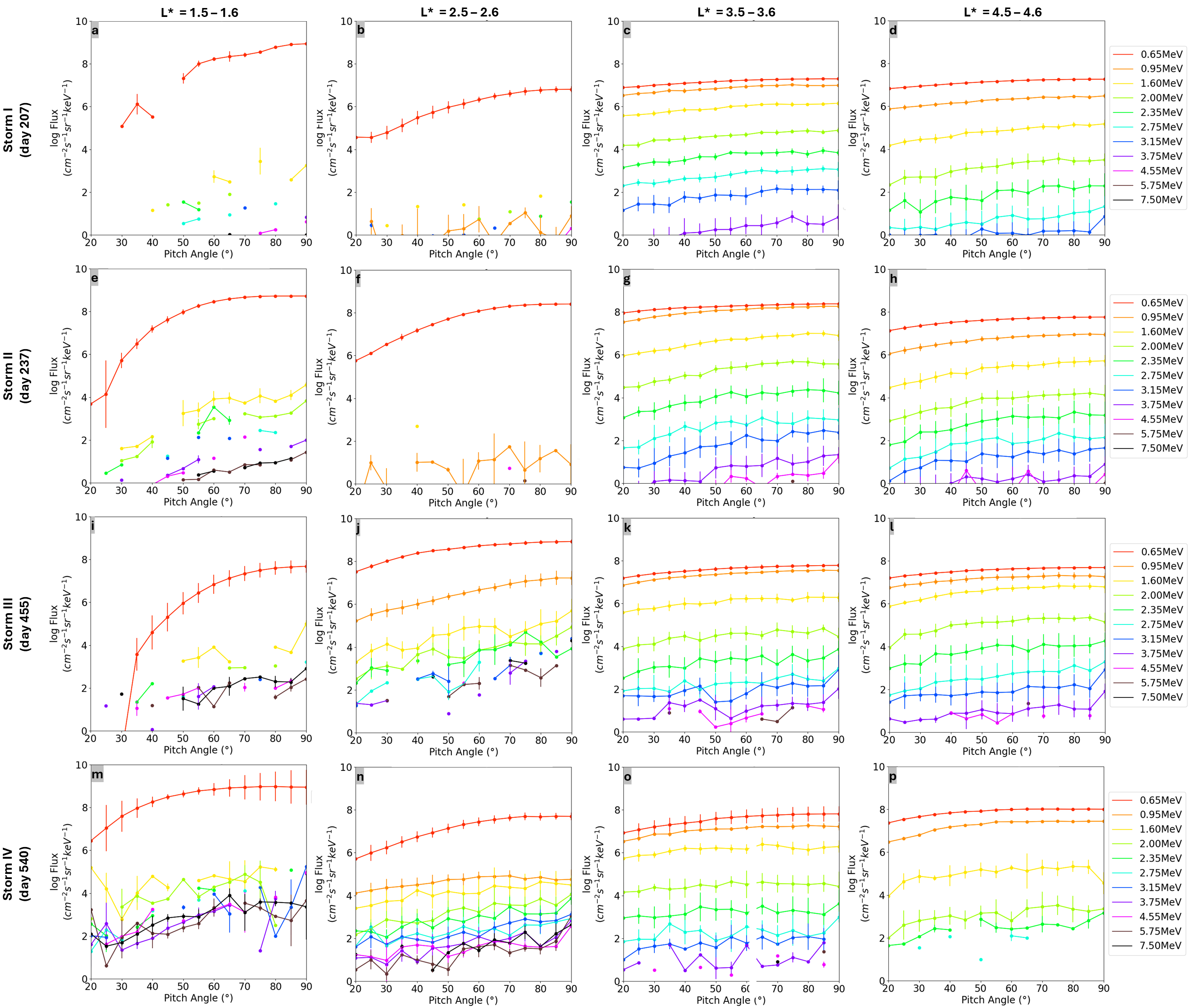}
     \end{center}
     
     \vspace{-5mm}
     \caption{Pitch angle distributions at different L* shells, used as a measure of identifying coherent or contaminated electron spectra. Day 207 = 26 Jul 1990; day 237 = 25 Aug 1990; day 455 = 31 Mar 1991; day 540 = 24 Jun 1991. }
      \label{PAspectra}
 \end{figure}

\section{HEEF Dataset}
\label{Section3}
\subsection{Instrumentation}

The data used in this study is derived from the Combined Release and Radiation Effects Satellite (CRRES) High Energy Electron Fluxometer (HEEF) instrument which obtained measurements in an equatorial geostationary transfer orbit moving through radial distances of 1.05-6.26 Earth radii at an inclination of 18$^\circ$, from 25 July 1990 till 12 October 1991 \citep{Gussenhoven96}.
The HEEF instrument \citep{Dichter93} was designed to measure the electrons in ten channels from 1 to 10 MeV with 10 degree half-angle pitch angle spectra obtained through the spinning of the platform and the CRRES magnetometer. The instrument utilised two solid state detectors and a bismuth germinate crystal scintilator and an anti-coincidence plastic scintillator, to correlate a particle detection and determine energy and species.
The HEEF instrument was extensively calibrated prior to launch \citep{Dichter92} however, shortly after launch it became necessary to turn off a heater in the HEEF compartment with the result that HEEF operated at temperatures significantly different than planned. Since the operation of the bismuth germinate (BGO) crystal scintillator is temperature sensitive, further calibration work on HEEF was completed using on-orbit data and laboratory calibration of a flight spare unit \citep{Hanser95}. Despite this the complete unfolding and convergence of the in-flight dataset was never accomplished with potential errors highlighted within the flux values \citep{McKellar96}. 

\begin{table}[ht]
\centering
\caption{Mid-point and ranges of the HEEF energy channels used in this study \citep{Brautigam95,McKellar96}.}
\begin{tabular}{|l|l|l|l|l|l|l|l|l|l|l|l|}
\hline
\textbf{\begin{tabular}[c]{@{}l@{}}Channel\end{tabular}} & 0 & 1 & 2 & 3 & 4 & 5 & 6 & 7 & 8 & 9 & 10 \\ \hline
\textbf{\begin{tabular}[c]{@{}l@{}}E$_{lower}$~\end{tabular}}    & 0.50 & 0.85 & 1.25 & 1.70 & 2.10 & 2.50 & 2.90 & 3.30 & 4.10 & 4.95 & 6.66 \\ \hline
\textbf{\begin{tabular}[c]{@{}l@{}}E$_{mid}$~\end{tabular}}      & 0.65 & 0.95 & 1.60 & 2.00 & 2.35 & 2.75 & 3.15 & 3.75 & 4.55 & 5.75 & 7.50 \\ \hline
\textbf{\begin{tabular}[c]{@{}l@{}}E$_{upper}$~\end{tabular}}    & $<$0.80 & $<$1.05 & $<$1.70 & $<$2.10 & $<$2.50 & $<$2.90 & $<$3.30 & $<$4.10 & $<$4.95 & $<$6.6 & $<$8.55 \\ \hline
\end{tabular}
\label{table}
\end{table}

In 2014, further empirical calibration factors were applied to the HEEF data to re-establish the utility of this dataset \citep{Johnston14}. 
Starting from a data set version with temperature and dead-time corrections applied, further corrections were implemented, including removal of data with incomplete pitch angle distributions and cross-calibration with the CRRES Medium Electron Analyser (MEA) \citep{Vampola92}. 
The resultant dataset consists of one-minute averages of differential electron fluxes separated into 5$^\circ$ pitch angle bins, with L* values derived from the \citet{Olson} magnetic field model.
Table \ref{table} outlines ten differential energy channels, eight of which are intrinsic to the instrument mode of operation and two additional differential channels (0.65 and 0.95 MeV) derived from differencing pairs of integral channels \citep{Brautigam95}. For further description of the cleaning of the data set and the contents, the reader is referred to \citet{Johnston14}.

\subsection{Deriving Empirical Loss Timescales}

Empirical electron lifetimes can be calculated using the same solution to Equation 1 through fitting the timescale, $\tau$, in Equation 2 directly to the HEEF data. We use a linear fit to the natural logarithm of the data to reduce interference from the largest fluxes.   The derivation of electron lifetimes utilises daily averaged electron fluxes and their temporal evolution within individual energy bins over extended periods. {The data does not consist of daily averages, rather the average of each pass through 0.1 L$^*$ shell ranges. {In the outer belt and slot region} the electron distribution functions, $A(\alpha_0)$, are summed between local pitch angles 60-120 degrees {to maximise counts} and taken for observations between $\pm$15 degrees magnetic latitude. {In the inner belt, 90 degrees pitch angle fluxes are used only, as the lifetimes here are highly pitch angles dependent}. The data is smoothed over a period of 1-2~days and decays fitted  using the exponential function over periods not less than 5~days, although the results are relatively insensitive to these specific parameters. }
To determine the quality of the fits we used the Pearson correlation coefficient, a value between -1 and 1, 0 being no correlation, -1 being a perfect negative correlation and 1 a perfect positive, with a threshold of --0.95 for determining an accurate timescale.
Following periods of exponential decay, the flux levels exhibit significant variations likely representing a balance of production and loss and radial transport and the assumptions behind using Equations \ref{decay} are no longer valid.

This process is automated with predefined criteria specifying the determination of a fit to a decay timescale. 
For automatic fitting, each flux data point is sequentially selected to be the starting point for a fit, with the initial end point determined by the given minimum fit range value. The resulting fit is used to find the Pearson coefficient and standard deviation.
If a fit is found, i.e. the algorithm returns a negative Pearson coefficient with a value less than the threshold, the end point is incremented, with this process continuing until the Pearson value is no longer lower than the threshold. This incrementation of the fit length is continued within a given tolerance of Pearson values to avoid local minima. 
If the fit returns a Pearson with a value greater than the threshold, both the start and end points are incremented and the fitting process is repeated on the new data range until the Pearson value becomes less than the threshold, i.e. a new decay is found. 

\begin{figure}[ht]
     \begin{center}
             \includegraphics[width=1.0\textwidth]{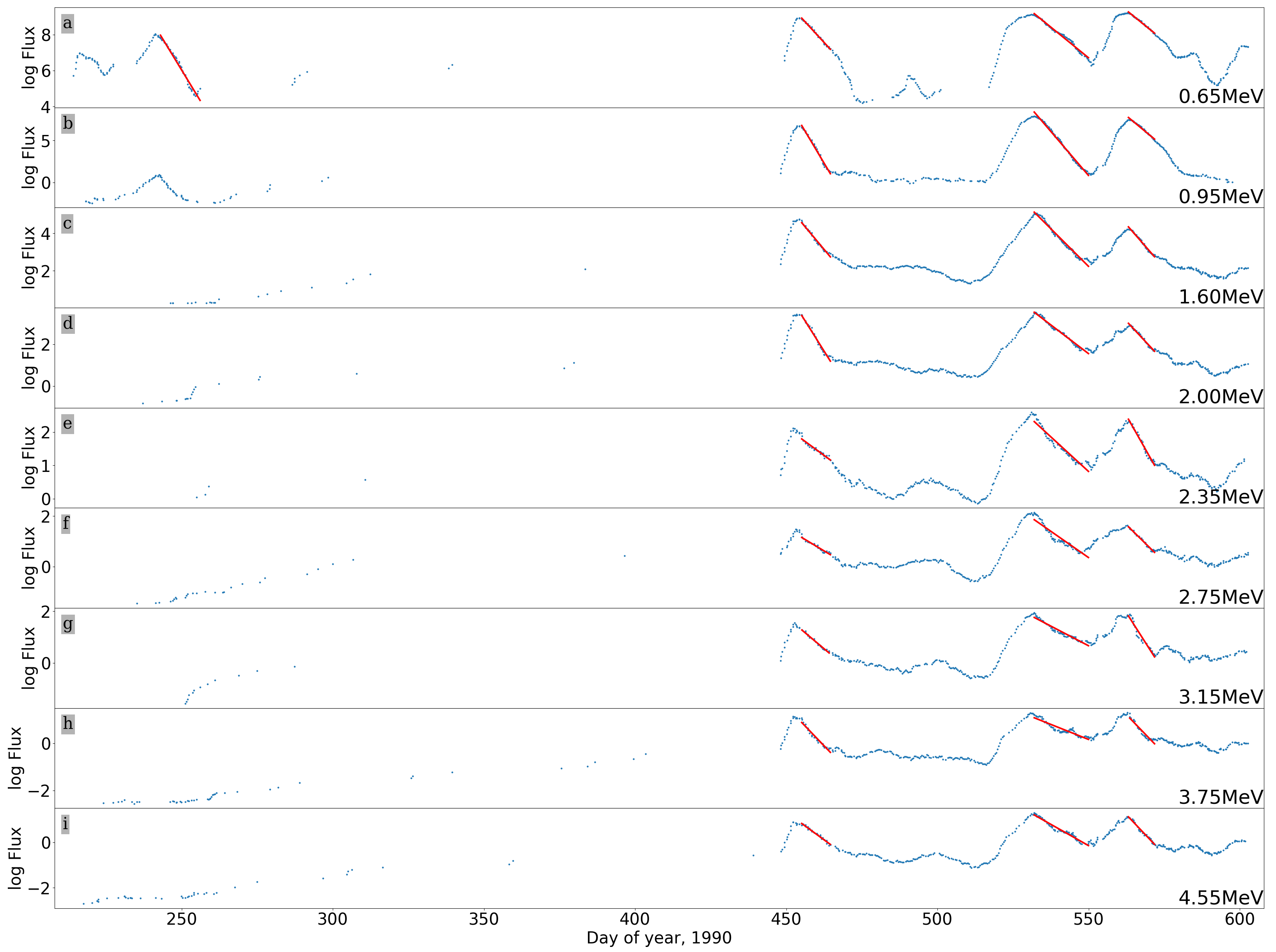}
     \end{center}
     \vspace{-5mm}
     \caption{Slot region electron differential number flux at L*=2.5 as a function of day number for the entire CRRES mission. (a-i) shows 0.65, 0.95, 1.60, 2.00, 2.35, 2.75, 3.15, 3.75, and 4.55 MeV electrons, within a L*=0.1 bin. Fitted exponential decays are overplotted in red for a minor storm at the storm of the mission and three severe storms during the latter half.}
      \label{fig3}
 \end{figure}

\section{Geomagnetic Storms in the CRRES Era}
\label{Section4}

Figure \ref{spectrogram} shows the HEEF data throughout the CRRES mission ordered by day numbers continuous from day 1=1 Jan 1990. Specifically, we show the averaged electron differential number flux as a function of L* and time at energies of  (a-f) 0.65, 1.60, 2.35,  3.15, 4.55, and 7.50 MeV. In the absence of continuous upstream solar wind measurements, panels (g) and (h), respectively, show a trace of the Kp and disturbance storm time (Dst) indices {as a measure of} the surrounding geomagnetic activity. The Kp indices are ordered according to the NOAA geomagnetic storm scales; Kp$<$5, Kp$\geq$5 (G1), Kp$\geq$6 (G2), Kp$\geq$7 (G3) and Kp$\geq$8 (G4), and the disturbance storm time (Dst) indices according to the storm definition of \citet{Loewe}; Dst$<$-30 nT (weak), Dst$<$-50 nT (moderate), Dst$<$-100 nT (strong) and Dst$<$–200 nT (severe). {As CRRES ended in October 1991 it did observe any} great/extreme events defined by Kp=9 (G5) or Dst$<$-350 nT, {although one did notably occur one month after the mission ended, on 9-10 November 1991, where Dst reached --354 nT \citep{Cliver09}} .  

Figure \ref{spectrogram} shows during the first half of the mission there were four strong storms and these are all associated with enhancements of 0.65 MeV electrons down to low values of L* with some events extending deep into the slot region and even the inner radiation belt. These storms tend also to be associated with flux enhancements at higher energies, up to 4.55 MeV. Interestingly, there is also a moderate storm that doesn’t quite reach the threshold to be classified as strong a few days before the second moderate storm and this is also associated with enhancements of 0.65 MeV electrons to low L*.  In sharp contrast there were twelve strong storms and two severe storms during the second half of the mission which result in a larger number of significant enhancements of 0.65 MeV electrons down to low L* and larger flux enhancements at higher energies.

The Kp index notably shows different scalings to the Dst in that in the second half of the mission three storms reach Kp=9- (G4), an equivalence between them in contrast to the Dst index showing the single event on March 1991 reaching an intensity significantly greater than the others. The Kp index also reveals several peaks within each of the three G4 storms, for example in March 1991 showing two Kp=9- peaks corresponding to a first large compression reported early on 24 March \citep{Blake1992} where the magnetopause was pushed inside geosychronous orbit for over five hours \citep{Elphic91} and then a further large compression late on 24 March and 25 March, reported to drive large magnetopause oscillations near geosychronous orbit \citep{Cahill92,Desai21}. The Kp index is a mid-latitude metric and this could be caused by motion of the auroral oval due to dynamic pressure enhancements associated with successive coronal mass ejections \citep{Koehn23}. The Dst index is conversely an equatorial index, associated with southward directed interplanetary magnetic fields and build-up of the ring current, and this therefore captures the geomagnetic response across longer time-periods. Several studies have highlighted that radiation belt flux variations may be better correlated with the Kp index \citep{Borovsky17,Wang23}.

{ Significant attention has been devoted to deriving the properties of the interplanetary shock that struck on 24 March 1991 (day 447) due to the rapid nature that the slot region was filled with indications that it reached 1500 km/s at the Earth \citep{Blake1992,Li1993,Hudson97,Elkington04}. 
{The June storm (day 524) and subsequence disturbances persisted for nearly 20 days, and featured four geomagnetic storms in total of varying intensity, the magnetospheric responses identified as likely reflecting successive high-speed solar wind structures and periods of strong southward solar wind magnetic field \citep{Gao1997}.}
Although the solar wind conditions have not been analysed, the July 1991 (day 554) storm was similarly inferred to derive from a large enhancement in dynamic pressure which pushed the magnetopause inside geosynchonous orbit for 3~hours, and large sudden storm commencement electric field that induced notable effects in the ionosphere \citep{Burke00,Wilson01}. The storm of 04 June 1991 (day 520) received less attention, but is listed 57 in the most extreme storms between 1868 and 2010, as ordered by the AA index \citep{vennerstrom2016}. This is two places above the 24 March 1991 event which sits at 59 and the July event is not listed. This AA-ordering further highlights inconsistencies between different geomagnetic indices. The lack of continuous upstream solar wind observation in 1991, with only intermittent observations from IMP8 when orbiting in the solar wind, unfortunately precludes further analysis of the upstream drivers of these storms observed by CRRES. }

\begin{figure}[ht]
     \begin{center}
             \includegraphics[width=0.6\textwidth]{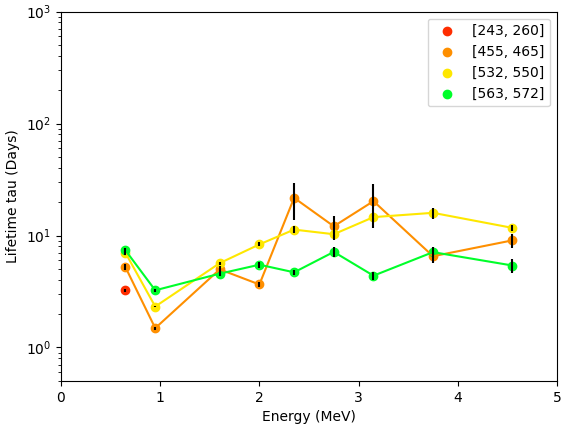}
     \end{center}
     \vspace{-5mm}
     \caption{Electron loss timescales in the slot region at L* = 2.5-2.6, as a function of energy for four doy time-periods during the CRRES mission noted in the legend.}
      \label{fig4}
 \end{figure}
\begin{figure}[ht]
     \begin{center}
             \includegraphics[width=1.0\textwidth]{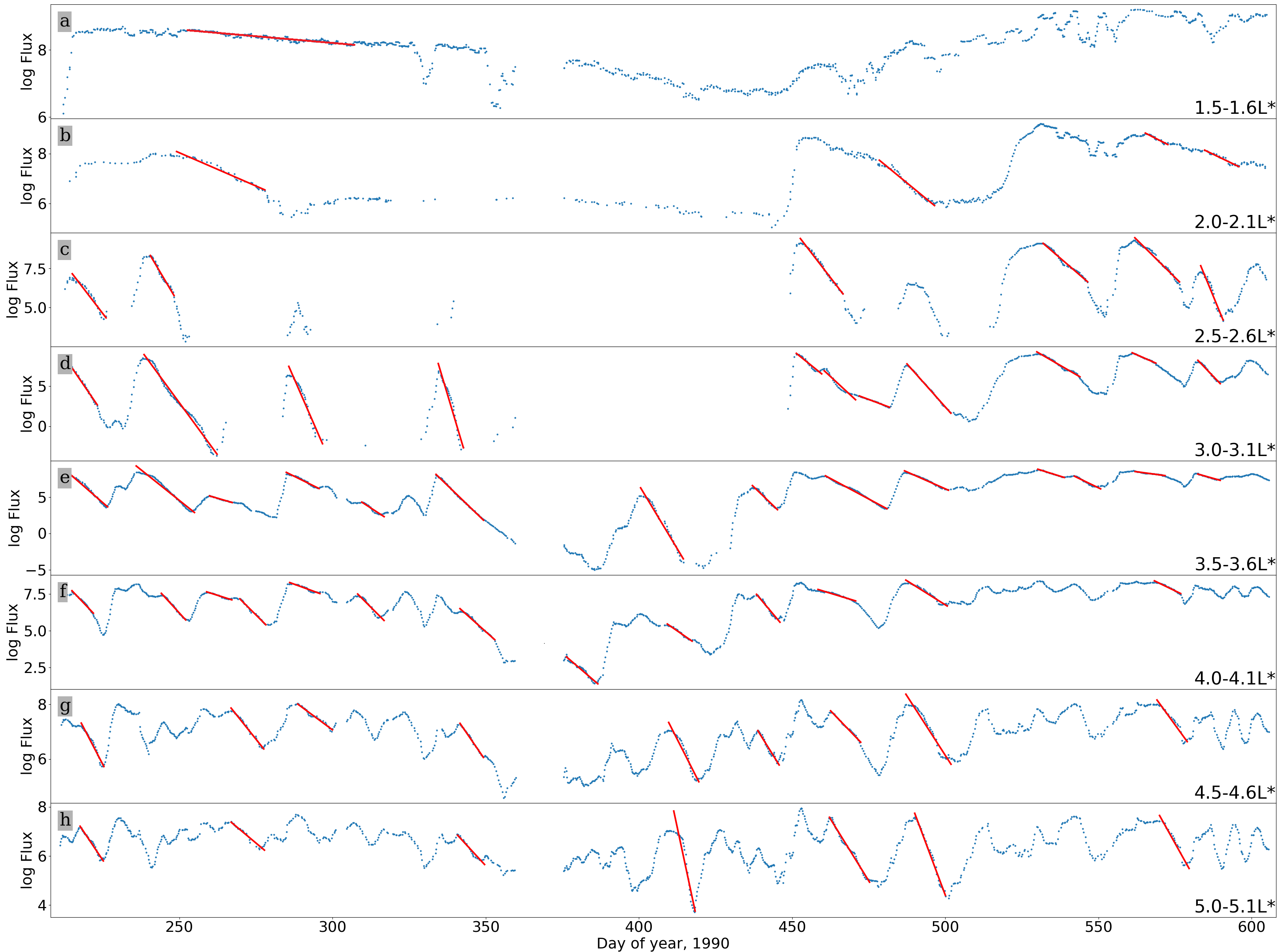}
     \end{center}
     \caption{The 0.65 MeV electron energy number flux as a function of day number for the entire CRRES mission for (a-g) 1.5-1.6L*, 2.5-2.6L*, 3.0-3.1L*, 3.5-3.6L*, 4.0-4.1L*, 4.5-4.6L* and 5.0-5.1L*. The fitted exponential decays are overplotted in red}
      \label{0.65MeV}
 \end{figure}

The outer radiation belt appears highly variable throughout the CRRES mission driven by elevated geomagnetic activity. Following the geomagnetic storm of 24 March 1991, the multi-MeV electrons in the outer belt are separated into two further belts as examined by \citet{Kellerman14} at lower energies. The corresponding energy and PA spectra are coherent up to 4.55 MeV and this multi-belt structure therefore also manifests at multi-MeV energies.
The flux levels at low counts appear highly variable, and we thus determine an approximate ``noise floor" in the data below which we do not try to interpret signals as genuine. This high noise floor, unfortunately, precludes the analysis in the study of the two highest energy channels 9 and 10, 5.75 and 7.5 MeV respectively, in this study.

Figure \ref{PAspectra} shows HEEF energy and pitch angle (PA) distributions corresponding to the data in Figure \ref{spectrogram} during three large geomagnetic storms and a lesser fourth storm on 26 August 1991, day 238. For each of these, the spectra are displayed at four distinct radial distances, in the inner belt at L* $ = 1.5$, in the heart of the slot region at L* $= 2.5$ and then in the outer belt at L* $ = 3.5$ and L* $ = 4.5$. 
In Figures \ref{spectrogram} and \ref{PAspectra}, an inner electron radiation belt is visible at L*$<$2 for the whole mission. In the lowest-energy channel 0, centred at 0.65 MeV, the dynamics of this inner belt correlate well with enhanced fluxes at higher L shells. The PA spectra at energy channel 0 (0.65 MeV) MeV at L* = 1.5 in Figure \ref{PAspectra} also agrees well with expectations appearing symmetric about 90${^\circ}$. The next energy channel 1 (0.95 MeV) is, however, devoid of fluxes in the inner belt. The higher energy channels do display counts here, but the non-uniform spectra are suggestive that this is proton contamination at multi-MeV energies in the inner zone. We are therefore unable to ascertain whether MeV and multi-MeV electrons appeared within the inner belt during the CRRES mission. 

Within the slot region, see L*$ = 2.5$ panel in Figure \ref{PAspectra} and Figures 1a-f,  elevated flux levels start after 24 March 1991 (day 448) and extend right through to the end of the mission. 
The fluxes for the few orbits surrounding the injection of this new belt on day 448 \citep{Blake1992} are not present in the dataset due to the cleaning procedures applied \citep{Johnston14} but the decay of this belt is visible following day 450. The corresponding energy and PA spectra in Figure \ref{PAspectra}j show coherent observations up to several MeV inside the slot region and throughout the outer belt. The further geomagnetic storms after days 524 and 554 in Figure \ref{spectrogram} also produce elevated fluxes in the slot region of a similar or greater magnitude to those deriving from the 24 March 1991 event, with fluxes comparable to those in the outer belt. {The extended June 1991 disturbances \citep{Gao1997} correlates with an increase in the slot region electron flux over an extended time period, in contrast to the March 1991 more sudden increase. The July increase simillarly occurs over an extended period.}

The Kp reaches similar magnitudes for these events, but the Dst index is significantly lower. These events are notable in that during solar cycle 24, multi-MeV electrons were unable to penetrate below L*=2.8 \citep{Baker14}, with great solar activity in 2003 \citep{Baker04} and 2024 \citep{Pierrard24,Li25} required to overcome this. These events enable us to calculate slot-region loss timescales at multi-MeV energies, something that was not possible during the Van Allen probe era.

To examine the persistence of the multi-MeV electrons within these slot-filling events, Figure \ref{fig3} shows the data within an L*=0.1 bin, centred at L* = 2.5, in each energy channel up to 4.55 MeV.The first storm only displays data at the lowest energy channels but three decay periods following storms in the latter part of the mission are fitted to at all energy channels, overlaid as red lines, with Pearson correlation coefficients all lower than --0.95.
Figure \ref{fig4} shows the resultant lifetimes as a function of energy from these periods. 
The trends show increased lifetimes at sub-MeV energies which decreases to a minimum near 1 MeV associated with increased pitch angle diffusion into the loss cone \citep[][see Fig 3 therein]{Glauert24}. The lifetimes then increase again and level out at multi-MeV energies.

\begin{figure}[h]
     \begin{center}
             \includegraphics[width=0.9\textwidth]
             {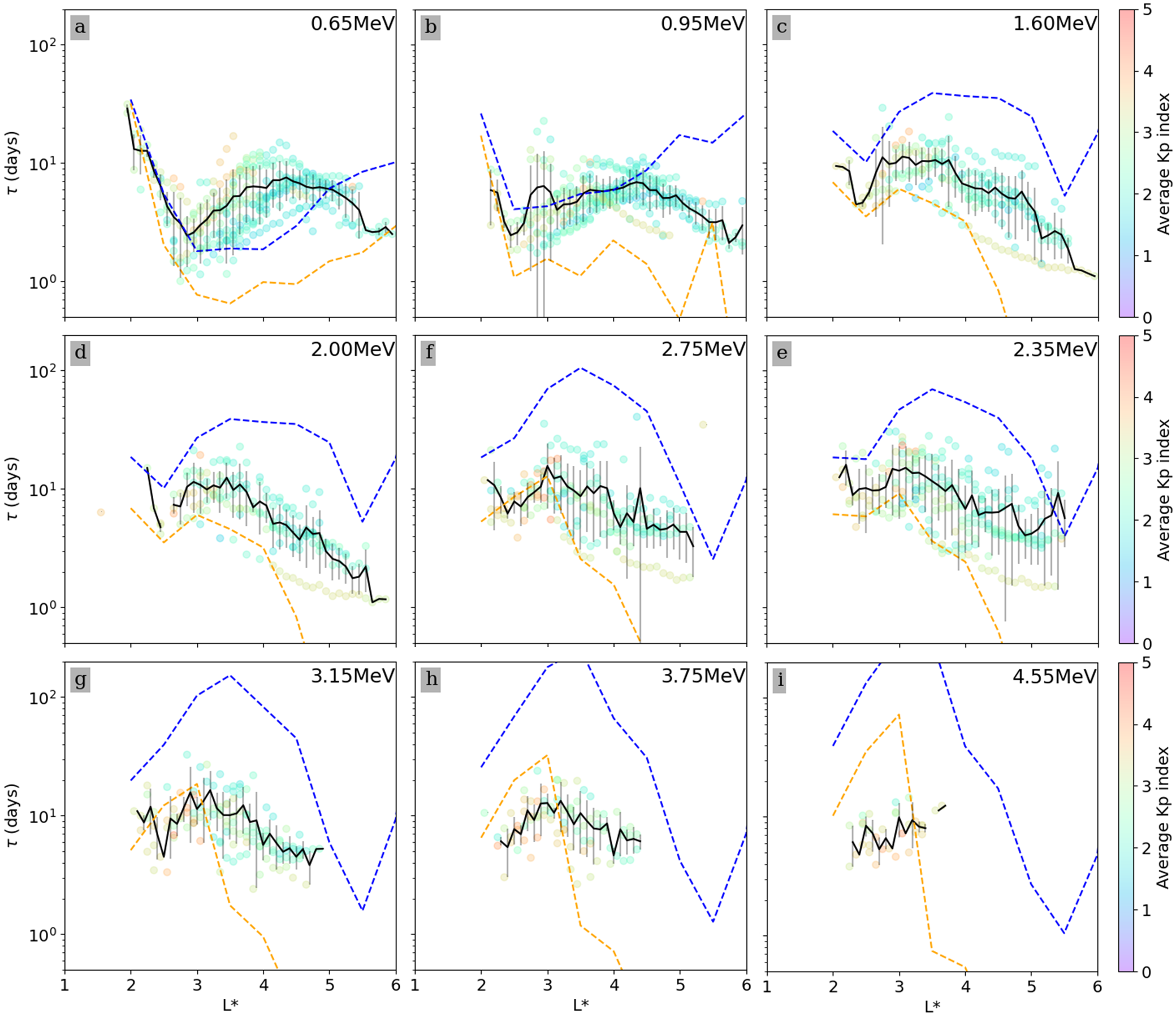}
     \end{center}
     \caption{HEEF {outer belt and slot region} lifetimes of multi-MeV electrons presented for each energy bin as a function of L*. The raw data are coloured according to the average Kp index during the {decay period}, the black line shows the average fit with error bars showing the standard deviation of calculated lifetimes. Pitch angle diffusion theoretical predictions are shown at Kp = 0-1 and 4-5.}
      \label{Kptrends}
 \end{figure}

\section{HEEF Electron Lifetimes}

To examine electron lifetimes throughout the CRRES era, Figure \ref{0.65MeV} then shows data across a range of L shells in the lowest energy channel, 0.65 MeV. The fits are once again overlaid on top of the data.
The automated algorithm returns increasing numbers of fits at the intermediate L shells examined, overall visibly well-capturing exponentially decaying fluxes. A few flux decreases can however also be seen to be missed. 
In some instances, e.g. Figure \ref{0.65MeV}d, it might appear that there are two stages to the decay process, for example in Figure \ref{0.65MeV}g after day 450. These often appear as a shallow gradient followed by a greater decrease in the differential number flux. 
Further non-uniform decays are evident near the peak fluxes themselves over shorter time periods, with notably rounded structures at the lower L shells. This may in part be due to increasing production close to the initial event, varying geomagnetic activity levels during the decay, or potentially different eigenmodes dominating the pitch angle diffusion.
At the lowest L shells, the electron fluxes decay over extended periods of tens-hundreds of days. The automated nature of the algorithm struggles with this extended period due to a shorter minimum fit window required to be defined to resolve the shorter timescales at larger L shells. {The fluxes also exhibits significant short-term variability in the belt, on the order of several days. While the pitch angle spectra in Figure \ref{PAspectra} appear coherent, {we restrict our lifetime calculations in this region to a single event between day 265 and 310 at 0.65 MeV}}.

The decay periods are well-represented by the exponential fits in the region 3.0 $\leq$ L* $\leq$ 4.5. At L* $\geq$ 5, only a few fits are visible. This may in part be due to the algorithm being difficult to optimise for such a wide spread in lifetimes where, at these greater radial distances, further transport processes associated with radial diffusion \citep{Su10}, drift shell splitting \citep{Sibeck87} and drift orbit bifurcations \citep{Desai21c} become significant.

Figure \ref{Kptrends} displays the results across all L$^*$ shells in the slot region and outer belt in each energy channel with each fit colored according to its mean Kp index {across the entire decay period. This allows analysis of the conditions that exist during the quiet time following the event and determine the electron lifetimes}. The average fit at each 0.1 L$^*$ bin is also shown as a black line together with error bars representing the standard deviation. 
At the lowest L$^*$ shells at the edge of the inner belt, the lifetimes of the 0.65 MeV electrons rapidly increase upward. {The inner belt lifetimes are not discussed further here with regards to geomagnetic activity, due to their pitch angle dependence, but are discussed in the subsequent section.}
Decreased slot region lifetimes between L* = 2-3 appear at this energy, as in Figure \ref{fig3}, before the lifetimes increase to another smaller maximum near L$^*$ = 4. 

The lifetimes of the 0.95 MeV electrons are lower than those of the 0.65 MeV electrons near L* = 4.0 but this trend does not extend into the slot region where the reverse is true.
There is a further peak near L$^*$ = 2.5-3 but this is due to a large spread in the data here. 
At energies greater than 1 MeV, the peak lifetime in the outer belt moves inward with increasing energy from L* $\approx$ 4.5 at 1 MeV down to L* $\approx$ 3 at 4.55 MeV. The location of the shortest lifetimes in the slot region fluxes also move inwards, with increasing energies displaying longer lifetimes. 

In addition to the average trends, the Kp values of each fitted lifetime indicate how the decays scale with geomagnetic activity. At higher L$^*$ shells, i.e. greater than L* = 3, elevated geomagnetic activity appears to drive fast losses with some of the shortest timescales showing Kp values near 5. At the lower L$^*$'s, however, and particularly at the lowest energy channel of 0.65 MeV, high Kp values often correspond to longer decay timescales. High Kp scenarios will still result in injections from the tail and subsequent acceleration of energetic particles and this trend therefore might be explained by sustained production of electrons occurring alongside pitch angle scattering, a phenomenon which preferentially affects lower energies. This is further discussed in Section 5.

\begin{figure}[ht]
     \begin{center}
             \includegraphics[width=1.0\textwidth]
             {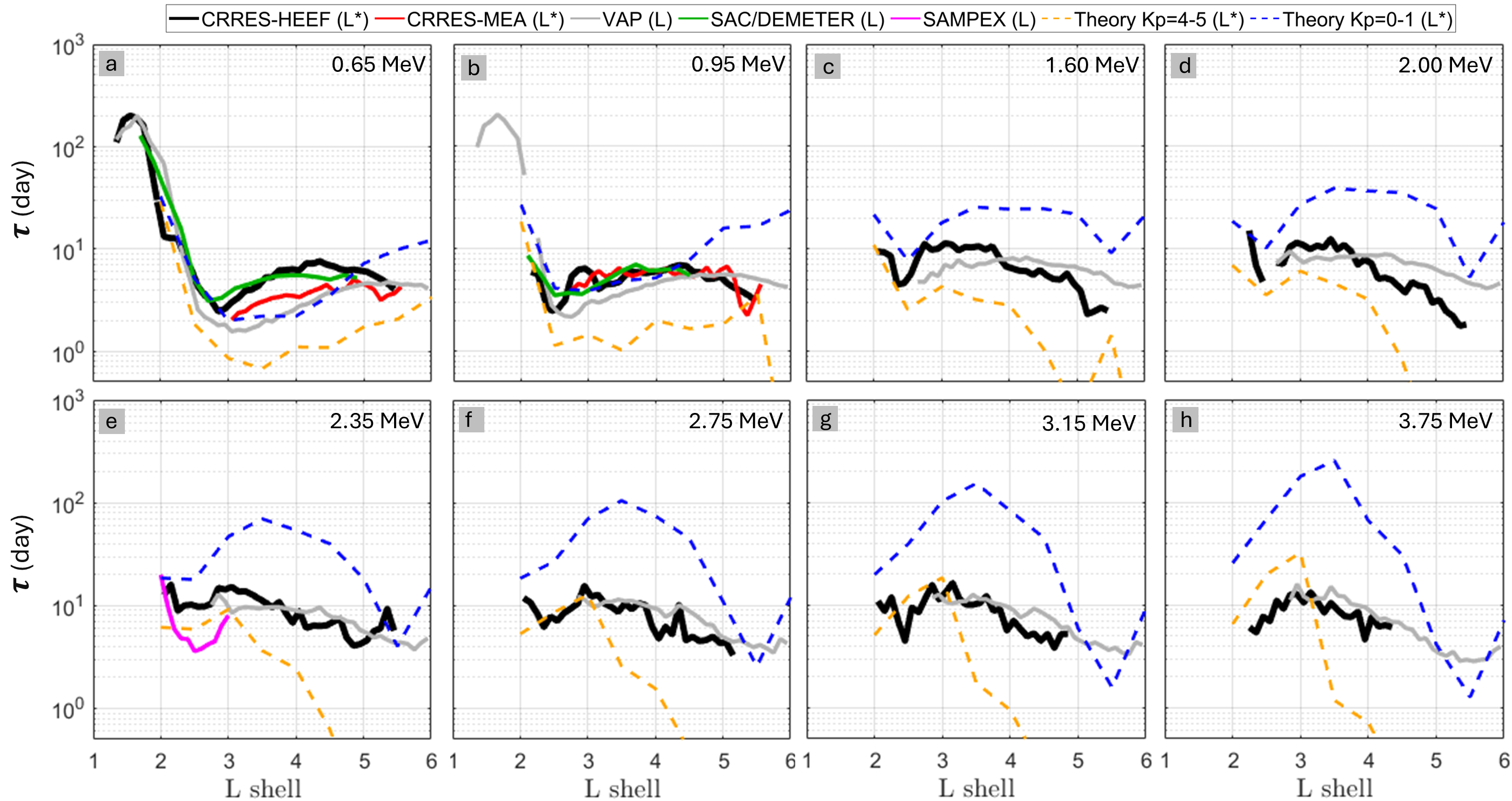}
     \end{center}
     \caption{
     Experimental and theoretical electron lifetimes as a function of L shell in the outer belt and slot region for (a-h) 0.65, 0.95, 1.60, 2.00, 2.35, 2.75, 3.15 and 3.75 MeV electrons respectively. The black, red, grey, green and pink traces represent the experimental lifetimes derived from the CRRES-HEEF (this study), CRRES-MEA also from cycle 22 \citep{Meredith06}, VAP from cycle 24 \citep{Claudepierre20a}, and SAC/DEMETER \citep{Benck10} and SAMPEX \citep{Meredith09} datasets from cycle 23. The dashed blue and orange traces represent the pitch angle diffusion theoretical lifetimes for quiet and high geomagnetic activity, respectively. All data is interpolated to the HEEF energies, whereas the SAMPEX 2--6 MeV data are associated with the 2.35 MeV HEEF channel.}
      \label{comparisons}
 \end{figure}

\section{Comparisons \& Discussion}
\label{Section5}
To compare the HEEF results with pitch angle diffusion theory, the pitch angle diffusion lifetimes described in Section \ref{Section2} are also plotted in Figure \ref{Kptrends} with dashed lines representing decays for minimum and maximum Kp levels of 0-1 and 4-5 respectively.  
The average HEEF results are also plotted in Figure \ref{comparisons} with further empirical results from other spacecraft. These include coinciding measurements from the CRRES-Medium Energy Analyser (CRRES-MEA) \citep{Meredith06}, from VAP between 2012 and 2019 \citep{Claudepierre20a}, from ACE/DEMETER from year 2000 \citep{Benck10} and from SAMPEX in 2003 \citep{Meredith09}.  All the lifetimes are interpolated to the HEEF energy channels with the 2--6 MeV SAMPEX data judged to best represent 2.35 MeV, see for example the relative fluxes in Figure \ref{PAspectra}. The CRRES and theoretical results are expressed in terms of L* whereas the further empirical results are expressed in terms of dipole L shell. The subsequent discussion utilises the term L shell when comparisons are between lifetimes utilising these different measures of the radial location of electron orbits.

In Figure \ref{Kptrends}, the pitch angle diffusion theory matches the observed trend of increased lifetimes near the inner belt at 0.65 MeV and depressed lifetimes in the slot region. 
In the inner part of the outer radiation belt, the quiet-time theoretical lifetimes at this energy are a factor of two or more smaller than the observed lifetimes in the region 3.2 $< L* <$ 4.6. These larger HEEF lifetimes are similar to those observed by SAC/DEMETER in cycle 23 \citep{Benck10} and also by ATS-1 \citep{Vampola71} in cycle 20 see \citet[][Fig. 2f therein]{Claudepierre20a}. This however is not observed in the VAP lifetimes. Further examination of Figure \ref{Kptrends} reveals a cause of this discrepancy. The underlying HEEF fits, most visibly between L* = 3-4, feature a clustering of lifetimes at the lower lifetimes ranges, similar to those reported by the MEA and closer to those observed by VAP. However, an additional clustering above this demonstrates a bi-modality within the dataset, with this second cluster corresponding to higher Kp values. This bimodality is also identifiable in the underlying fits on Figure \ref{0.65MeV}e,f where many of the fitted lifetimes have shallower gradients.  

The consequent differences in the average lifetimes in Figure \ref{comparisons}a may therefore have been caused by differing levels of geomagnetic activity,
with CRRES having observed an average Kp index of 2.67 compared to the VAP average of 1.67. This can be explained through three distinct mechanisms. Firstly, the intensity of geomagnetic storms were significantly greater in solar cycle 22 compared with solar cycle 24. These, during the latter half of the mission, indeed appear to be associated with the slower decays in Figure \ref{0.65MeV}. These intense and longer lasting events influence the average Kp indices of a decay and thus skew the decay rates to higher Kp values, producing a ``left-right" Kp bias as indeed shown by \citet[][Figure 6 therein]{Meredith06}. Secondly, disturbed geomagnetic conditions would more frequently reoccur during a decay period. This is visibly consistent with several of the decays observed, see for example the decays days 270 and 295 in Figure \ref{0.65MeV}f-g where large injections/enhancements occur during the decay period. Thirdly, it is well-known that typical solar wind parameters have declined significantly over the past 50 years, with differences between the 1970-1990's and 2010's having been highlighted as high as tens to over fifty percent \citep{McComas13}. Such long-term trends in the radiation belts are difficult to study due to their inherent variability, and the inconsistency of dedicated missions, but the longitudinal comparison between CRRES and VAP data, may enable such trends to be identified. Higher ambient solar wind driving might therefore have resulted in enhanced background levels of magnetotail loading and substorms injections. In addition, each of the above three points might also cause the plasmasphere/plasmaspheric hiss to extend to lower L shells during the CRRES era than during the VAP era. Lower energies are particularly affected by this loss mechanism, as found by the evolution of the energy spectra within VAP and CRRES observations \citep{Zhao19,Johnston10}.

The bimodality in the decays rates is less apparent in the lifetimes beyond L* = 4, with the high Kp lifetimes now more often lower than the low Kp lifetimes and the average even lower than the VAP observations beyond L* = 5.  The CRRES-MEA \citep{Meredith06} lifetimes are lower than the HEEF lifetimes in the outer belt between L = 2-4 although are higher than determined by VAP. The reason behind the differences with the MEA dataset is initially surprising given the HEEF and MEA datasets have been cross-calibrated \citep{Johnston14}. Comparisons between the \citet{Meredith06} fits at the MEA energy channel of 0.604 MeV (not shown) and the HEEF energy channel of 0.65 MeV in Figure \ref{0.65MeV}, indicate that the HEEF fitting algorithm skews towards longer fits. This is evident near days 270, 495 and 520 in Figure \ref{0.65MeV}f as these events feature some intrinsic variability in the fluxes, and Kp and DST indices, see Figure \ref{spectrogram}, but which overall are still well-approximated by periods of exponential decay if the algorithm is able to move past local minima in the Pearson values. This type of discrepancy may also explain further differences between the various studies reporting electron loss timescales. 

A further factor to be considered is that \citet{Brautigam95} adopt an energy span of 0.50-0.80 MeV for HEEF energy channel 0 and the electron lifetimes derived from HEEF for 0.65 MeV are therefore influenced by longer lifetime electrons than the MEA lifetimes. This is the energy-dependent decay seen in \citet{Meredith06} and \citet{Zhao19}, i.e. as lower energy electrons are depleted the higher energy ones will dominate and may thus influence the lifetime more.

It should be noted that \citet{Claudepierre20a} identify contamination due to Brehmstrahlung from higher energy electrons as a further potential cause of these longer lifetimes in the MEA and SAC/DEMETER datasets. As CRRES did not downlink the data required to identify and correct this effect, it is not possible to ascertain the extent to which HEEF was affected via the same method as VAP. 
However, we don’t tend to observe features at 0.65 MeV similar to those reported by \citet{Claudepierre20a} at 0.6 MeV in the VAP uncorrected data, which tend to show two different decay periods when the contamination is present, with a rapid initial decay and then a more gradual decay. This suggests that contamination from Bremsstrahlung is not so significant for the HEEF data.
The 0.65 MeV lifetimes also peak beyond L* = 4, whereas the multi-MeV lifetimes peak below L* = 3 and so the multi-MeV electrons are not directly correlated with the elevated lifetimes at 0.65 MeV. The HEEF and VAP lifetimes at higher energies also show good agreement, see subsequent discussion, and the HEEF lifetimes are actually shorter in the outer belt, consistent with higher accompanying geomagnetic activity. 

In the inner belt, the decay timescales match closely with the VAP decay timescales, which were also determined for 90 degrees flux.
These elevated fluxes in the HEEF dataset, see Figure \ref{0.65MeV}a, appear to have been caused by storms at or prior to the start of the mission followed by relatively quiet conditions which allow their decay timescales to be calculated across this period. 
These loss timescales approach 200 days and is also evident in that during the second half of the mission, see again Figure \ref{0.65MeV}a where geomagnetic activity was higher, the 0.65 MeV flux is unable to decay and appears to be largely on an upward trend from around day 450 to ~550. This points to the controlling influence of slot region fluxes on electron flux levels in the inner belt.

At 0.95 MeV, Figure \ref{Kptrends}b and \ref{comparisons}b, the theory bounds many of the fitted lifetimes but is lower than the observations when scaling by geomagnetic activity with half the fitted lifetimes outside the upper bound predicted by theory. The HEEF, MEA and SAC/DEMETER observations however show similar lifetimes at all L shells. The VAP lifetimes appear slightly lower, consistent again with the lower geomagnetic activity argument discussed previously with reference to Figure \ref{comparisons}a.
At 1.60 MeV, Figure \ref{Kptrends}c and \ref{comparisons}c, the HEEF and VAP observations show similar average values at L shells of 4.5-5.5 but further in, the HEEF results decay more slowly in-line with the previously discussed high Kp-associated decays. This trend is similar at 2.00 MeV, but the HEEF lifetimes are noticeably shorter than the VAP lifetimes beyond L shells of 4. The HEEF lifetimes are similar to the VAP lifetimes at 2.35, 3.75 and 3.15 MeV, but again do decay slightly faster at L shells greater than 4. The HEEF observations provide the first calculation of lifetimes at $>$4 MeV, and comparisons with previous observations at 4.55 MeV are therefore not possible. The fact that the VAP average lifetimes peak further out than the HEEF lifetimes may also in part be due to the VAP lifetimes using L instead of L* \citep[e.g.][]{Roederer18}. 

At energies of 1.6, 2, 2.3 and 2.75 MeV (Figure \ref{Kptrends}c-f), the theory captures the full spread in the HEEF data. At 2.75 MeV, Figure \ref{Kptrends}g, the theory still bounds the data at L shells greater than 3.5 but at lower radial distances and in the slot region diverges slightly with the prediction at Kp=4-5 appearing above many of the HEEF lifetimes. This trend is exaggerated further at 3.75 MeV and 4.55, Figure \ref{Kptrends}h-g, with the theory predicting longer lifetimes in the slot region than observed and a wide spread in possible lifetimes throughout the outer belt. At 4.55 MeV, the theory predicts even longer lifetimes than at 3.75 MeV, an increase not seen in the HEEF decay rates. 

At L shells less than 4, discrepancies between theory and observations have been highlighted to potentially result from the lack of EMIC wave models \citep{He22} which preferentially affect larger energies, lower-frequency hiss \citep{Ni14} or wave-normal angle effects \citep{Hartley18}. The EMIC wave models used herein indeed only extend down to L* = 3.25, below which these waves are difficult to identify. The discrepancies between the observations and the theory could also well-derive from the assumption in Equation 1 of pure pitch angle diffusion. For example, \citet{Ross21} employ 3-D modelling to better capture the decay of 2.5 MeV electrons near L = 3.5 using combined energy and radial diffusion. \citet{Su12} indeed highlight that at L shells greater than 5, the loss timescales are largely insensitive to energy due to radial transport. It is therefore likely that radial diffusion also plays a role at lower L shells and particularly during higher levels of geomagnetic activity when the magnetopause may be more compressed. {\citet{Broll23} also shows that simulated pitch angle diffusion evolves differently than empirical estimates when non-equilibrium pitch angle distributions are considered, {and indeed shows this approach yield decay timescales much closer to the observed timescales than the theoretical `lifetimes' (from $D_{\alpha\alpha}$'s slowest-decaying mode) in the inner belt}}.

 Although the HEEF lifetimes are well-bounded by the theory, the lifetimes are not strictly ordered by geomagnetic activity, with decays at lower L shells having been highlighted as corresponding to higher Kp indices. An event specific spread in the lifetimes at higher L shells is less apparent but some of the intermediate decays associated with Kp values of 1-3 are still not correctly ordering the decays. This suggests the event specific nature of the decays \citep{Ripoll16}, potential role of alternative transport processes such as energy and radial diffusion, {and errors with using a single global geomagnetic activity index for parametrising radiation belt dynamics}.

\section{Summary \& Conclusions}
\label{Section6}
This study has utilised the CRRES-HEEF instrument to analyse relativistic electrons during {the maximum of} solar cycle 22 with well-resolved measurements extending from 0.65 to 4.55 MeV. 
The slot region was observed to be flooded with electrons up to at least 4.55 MeV following two severe and one strong geomagnetic storms in the latter half of the mission. Comparable events have only been observed on a few occasions since, for example during and following the Hallowe'en storm period of 2003 in solar cycle 23 \citep{Meredith09,Baker13} and during the Mother's day/Gannon storm of May 2024 in solar cycle 25 \citep{Pierrard24,Li25}, with solar cycle 24 identified as possessing an ``impenetrable'' barrier preventing high energy electrons from reaching below L*= 2.8 \citep{Baker14}. 
 {The regularity of this phenomenon during the latter half of the mission can therefore only be interpreted as presenting the possibility of reoccurring during sustained periods of elevated solar activity, and therefore representing a direct risk to the increasing number of medium earth orbit (MEO) satellites operating in the slot region}. 

To examine the individual decay periods in more detail we  parameterise the results according to the wide spread in geomagnetic activity during this peiod. This reveals several important trends. At higher L shells, the results are generally well-ordered by geomagnetic activity with high Kp decays often driving shorter loss timescales. Near the inner regions of the outer belt and in the slot region, however, this did not appear to be the case with a series of slow decays corresponding to high levels of activity. These persistent trends suggest that pitch angle diffusion cannot solely explain the evolution of fluxes during this periods, with other competing transport processes, {such as energy and radial diffusion, likely contributing}. 
{The large geomagnetic storms which flooded the slot region, were also discussed in terms of their severity, with Kp, Dst and aa indices often diverging. While the Kp index has been highlighted as better predicting radiation belt dynamics \citep{Borovsky17}, the lack of strict ordering of electron lifetimes by this index indicates a more sophisticated metric may be better for parametrising electron lifetimes}.

The HEEF electron loss timescales were then compared to lifetimes  from other solar cycles and several differences were highlighted. These include longer-lasting sub-MeV electrons near the inner edge of the outer belt, the several slot-filling events which allowed the first definitive timescales at multi-MeV energies to be derived, and faster decaying fluxes beyond L* $\approx$ 5. These differences are associated with higher levels of geomagnetic activity {and match similarly elevated sub-MeV lifetimes in this region from solar cycle 23 and 20. This was therefore suggested to result from higher levels of solar and geomagnetic activity during {the maximum of} solar cycle 22 than VAP observations in cycle 24}.

A further outcome of this investigation is that this HEEF dataset provided valuable data regarding radiation belt dynamics during {the maximum of} solar cycle 22, either for stand-alone investigations or boundary conditions for radiation belt modelling. Care should be taken however, particularly when interpreting absolute fluxes, with sight of the underlying energy and pitch angle spectra. 

\section*{Acknowledgements}
J.P. and S.R. acknowledge University of Warwick Undergraduate Research Support Scheme (URSS) bursaries. RTD acknowledges a Science and Technology Facilities Council Ernest Rutherford Fellowship ST/W004801/1 and UKSA grant ST/Y005635/1. NPM and SAG were supported in part by Natural Environment Research Council grants NE/V00249X/1 (Sat-Risk), NE/X000389/1 and NE/R016038/1. The views expressed are those of the authors and do not necessarily reflect the official policy or position of the Department of the Air Force, the Department of Defense, or the U.S. government. We thank the reviewers and editor for their helpful comments on this manuscript.

\section*{Open Research Section}
The CRRES-HEEF dataset can be accessed on Zenodo at: \citet{johnston2014data}. The theoretical diffusion coefficients, loss timescales and pitch angle distributions reported can be accessed on the NERC EDS UK Polar Data Centre at \citet{glauert2024data}.

\bibliography{bibliography}

\end{document}